\documentclass[chicago]{emulateapj}
\usepackage{natbib}
\usepackage{graphicx}
\usepackage{times}
\usepackage{float}
\usepackage{rotating}
\usepackage{epstopdf}
\usepackage{multirow}
\usepackage{hyperref}
\usepackage{color}
\usepackage{natbib}
\usepackage{booktabs}
\usepackage{txfonts}
\usepackage{epsfig}

\shortauthors{Tavasoli et al.}

\include{def_bibtex}
\include{definitions}

\begin{document}

\title{The galaxy population in voids: Are all voids the same?}
\author{Tavasoli S.$^{1}$, Rahmani H.$^{1,2}$, Khosroshahi H. G.$^{1}$, Vasei K.$^{3}$, Lehnert M. D.$^{4}$}

\affiliation{$^{1}$School of Astronomy, Institute for Research in Fundamental Sciences (IPM), P.O. Box 19395-5531, Tehran, Iran}
\affiliation{$^{2}$Aix Marseille Université, CNRS, LAM (Laboratoire d'Astrophysique de Marseille) UMR 7326, 13388, Marseille, France}

\affiliation{$^{3}$Department of Physics and Astronomy, University of California, Riverside, CA 92521, USA}
\affiliation{$^{4}$Institut d$^{'}$Astrophysique de Paris, UMR 7095, CNRS,
Universit{\'e} Pierre et Marie Curie, 98 bis boulevard Arago, 75014
Paris, France}

\begin{abstract}
\noindent The influence of under-dense environments on the formation and
evolution of galaxies is studied by analysing the photometric
properties of $\sim$ 200 galaxies residing in voids, taken from our SDSS DR10 void
catalog up to z ${\sim}$ 0.055. We split void galaxies
into two subsamples based on the luminosity  density contrast of their host voids:
'sparse void' $\delta_s =\delta < -0.95$ and 'populous void' $\delta_p =\delta > -0.87$.
We find that galaxies in sparse voids are less massive than galaxies in populous voids. The 
luminosity distribution of galaxies in populous voids follows the same distribution  observed 
across the SDSS survey in the same redshift range. 
Galaxies in the sparse voids are also bluer suggesting that they may be going through a 
relatively slow and continuous star formation.  
Additionally, we find that the luminosity function of galaxies in populous voids is represented  with the Schechter function whereas
the same does not hold for sparse voids. 
Our analysis suggests that the properties of a host void
plays a significant role in the formation and evolution of the void galaxies 
and 
determining the large scale evolution of voids is an important step to  
 understand 
what processes regulate the evolution of galaxies. 
\end{abstract}


\section{Introduction}
\label{intro}

One of the main outstanding problems in observational cosmology is  to understand 
how galaxy properties are influenced by their environments and evolve with cosmic time.
For instance, in  over-dense regions 'groups/clusters', distinct mechanisms such as tidal force,
ram pressure stripping and harassment play a fundamental
role in galaxy  star formation rate, color and morphology 
\citep[e.g.][]{Veilleux2005,Kormendy2009}.
By incorporating these quenching mechanisms, galaxies in higher density regions tend to be redder
 and earlier type, have lower star formation rate and are more strongly clustered. Some of
these trends might lead to the well-known 'morphology-density' relation \citep[][]{Dressler1980}. 
In addition to these baryonic processes there are other mechanism that
can change the properties of the galaxies in different environments. The relation between dark matter perturbations in background and the distribution of dark matter halos (that host galaxies), which is known as the halo bias parameter, play a crucial role in the properties and mass distribution of the galaxies.
To understand the influence of environment on galaxy formation, most of the previous studies 
 have focused on the properties of galaxies in high-density regions \citep[e.g.][]{Scarlata2007,Bower2008}
 and few studies have focused on field and void galaxies
\citep[e.g.][]{Pustilnik2002,Rojas2004,Goldberg2005,Hoyle2005,Hoyle2012,Kreckel2012,Pan2012}.
 In this letter, we consider the other extreme case and study the influence of environment on galaxies which reside mainly in
the under-dense or void regions.
Since there are no complex processes such as close encounters and galaxy mergers in void regions,
void galaxies are excellent probes of the
effect of environment and cosmology on structure
formation and galaxy evolution

Early spectral and photometric studies of void galaxies have shown that they are statistically bluer, have a later 
morphological type and higher specific star formation rates than galaxies in average-density environments 
\citep[][]{Rojas2004,Rojas2005,Patiri2006}.
Recent, high-quality spectroscopic and photometric data from large
redshift surveys, and also modern N-body simulation can provide
valuable information on void regions \citep[][]{martel1990,weygaert1993,aragon2013,Jennings2013,Sutter2012,Sutter2014a,Tavasoli2013}.
The unique properties of void environments and their internal structures are appropriate tools for the study of cosmological models 
\citep[][]{Lavaux2010,Biswas2010,Ceccarelli2013},  putting constrains on cosmological parameters \citep{Betancort-Rijo2009}, testing theories of dark energies \citep[][]{Bos2012,Sutter2014b} and modified gravity \citep{Clampitt2013}.

In this study, we focus on the photometric properties of void galaxies in various under-dense regions drawn from Sloan Digital Sky Survey (SDSS DR10).
We use 1014 void galaxies which reside in 167 voids that are characterized by their luminosity density contrasts.

An important question is whether the formation of void galaxies is in anyway determined by the properties of
 the host voids. We attempt to address  this issue by separating the void galaxies located in more under-dense regions, which we refer to as  {\it sparse voids} from those that reside in denser regions referred  to  as {\it populous  voids}. 
We define  'sparse void' $\delta_s =\delta < -0.95$ and 'populous void' $\delta_p =\delta > -0.87$, the motivation for which is
 discussed in the Section 3.
We describe the observational data and sample selection in Section \ref{sample}. 
The properties of the void galaxies are discussed in Section 3. 
A summary and concluding remarks are presented in section 4.

Throughout this paper, we assume a flat $\Lambda$CDM cosmology and 
adopt following cosmological
parameters: the Hubble parameter H=70 km\,s$^{-1}$\,Mpc$^{-1}$ and the
matter density $\Omega_m=0.27$ \citep{Hinshaw2013}.

\begin{figure}
\centering
\hspace{-1.00cm}\includegraphics[width=0.40\textwidth,angle=90]{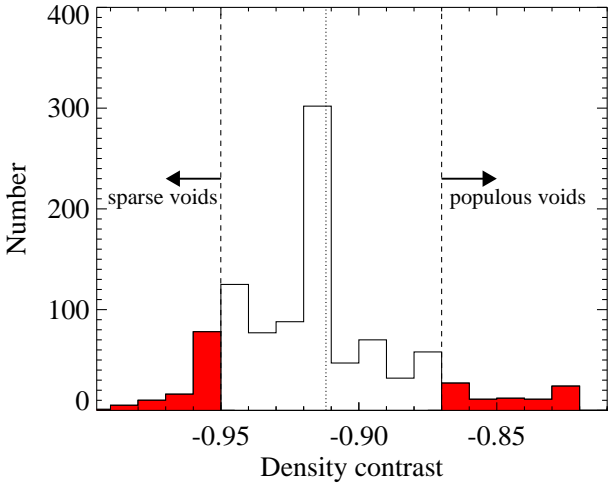}
\caption{The distribution of density contrast of void galaxies.
The dotted and dashed lines marks the mean and 1-$\sigma$ of the
distribution, respectively.
The two filled regions in the
left and right side of the histogram present 'sparse sample' and `populous sample', respectively.
There are 20 of void galaxies with contrast density higher than -0.81  
not shown  in this figure.
}
\label{contrast}
\end{figure}

\section{Sample selection}\label{sample}

To study the effect of under-dense environment on {formation and evolution}  of void galaxies,
we use a catalog of voids { extracted} from a volume-limited   
spectroscopic sample of SDSS DR10 \citep{Ahn2014} using the method    described in \citet{Tavasoli2013}.

The boundaries of the selected region of SDSS
are $135^\circ < {\rm RA} < 235^\circ$ and $0 <  {\rm DEC} < 55^\circ$ which contains ${\sim}$ 66000  galaxies
 with a limiting r-band magnitude of $m_{r, {\rm petrosian}} < 17.77$ up to z ${\sim}$ 0.055. 

The redshift of all selected galaxies are corrected for the motion
of the local group and are given in the CMB rest-frame. {
Furthermore}, the k-corrections of SDSS galaxies are carried out
using the \textsc{kcorrect}  algorithm developed by
\citet{Blanton2003} and \citet{Blanton2007}. In order to produce a
homogeneous sample of data suitable for the statistical study
of void galaxies, we take a volume-limited sample in the  redshift
range  $0.010 < {\rm z} < 0.055$. The upper limit for the redshift is defined 
by the limiting magnitude  ${M_r}$ = -19 and leaves ${\sim}$ 40000 galaxies in the final sample.
To extract a void
catalog from our SDSS spectroscopic sample, we apply the void finder algorithm introduced by \citealt{Aikio98}, which does not require voids to be
 spherical. 

Prior to applying this void finding algorithm, we classified wall and field galaxies based on the distance to the nearest neighbor \citep{Hoyle2002}. Whereas field galaxies are candidate as void galaxies, the AM algorithm starts on the cartesian gridded wall galaxy sample by defining a distance field. To assign each element in the grid sample to a subvoid, we employed the climbing algorithm \citep{schmidt2001}. Finally, if the distance between two subvoids is less than both distance fields, they will be joined into a larger void. The void volume was estimated using the number of grid points inside a given void multiplied by the volume associated with the grid cell.(see \citealt{Tavasoli2013} for further algorithm details).

The generated void catalog, includes variety of voids in size $R_v$, and
 luminosity density contrast $\delta_v$. The luminosity density contrast of a void is defined by
$\delta_v = (\rho_v-\rho_m)/\rho_m$ where $\rho_v$ is given by the ratio of the total luminosity
of galaxies inside a given void by the volume of that void and $\rho_m$ is mean luminosity density of the volume-limited sample. 
Hereafter, for simplicity, we use density contrast instead of luminosity density contrast. 
For each void, we defined its effective radius $R_v$ as the radius of a sphere whose volume is equal to that of the void.
In order to avoid counting spurious voids in our catalog, the size of voids should be larger than $R_v > 7$ Mpc.
{Our final catalog contains 167 voids within which 1014 void galaxies, brighter than -19,  reside.}

\section{Results}
In this section, we describe general properties of the void galaxies
that reside in various under-dense regions.The main aim is to
find a connection between the evolution of void galaxies and
density contrast of voids. { To characterize the environment of void
galaxies we attribute the density contrast of each void to all galaxies
residing in that void.} Fig. \ref{contrast} presents the
distribution of the density contrast associated with 1014 void galaxies
identified in 167 voids. 

The distribution has a mean of ${\approx}
-0.91$ with a standard deviation of 0.04 shown with dotted
and dashed lines, respectively. { Fig. \ref{contrast} shows that the under-dense
regions where void galaxies reside, have  different density
contrasts. 
 In order to { explore} the effect of
under-dense regions on the evolution of void galaxies, we 
define two subclasses of void galaxies according to the density
contrast of their host voids: 'sparse void' $\delta_s =\delta <
-0.95$ and 'populous void' $\delta_p =\delta > -0.87$. The two
classes are defined after rejecting all galaxies within $\pm
1\sigma$ around the mean contrast density. Hereafter we refer to
them as 's-sample' and 'p-sample' for simplicity, which represent
the void galaxies in sparse and populous  voids.There are 110 and
111 galaxies in our s-sample and p-sample located inside 38 and 25 voids, respectively. 
Based on the definition of void sphericity as given by \citet{Tavasoli2013}, the s- and p-voids have average sphericity of 0.71 and 0.69, respectively, 
with the standard deviation of 0.06 for both samples. Hence, there is no difference between the shape of the voids in two samples. 
However, the median size of the voids in s-sample, 13 Mpc $h^{-1}$, is ${\sim}$ 3 Mpc $h^{-1}$ larger than that of p-sample. 
The latter will effect the normalisation of the luminosity function (see Fig. \ref{lf}). 
In the { following subsections we compare  photometric properties (luminosity, color
and luminosity function) of void galaxies in s- and p-samples to trace the effect of various cosmic environment.}

\subsection{Luminosity}

\begin{figure}
\centering
\hspace{-1.00cm}\includegraphics[width=0.40\textwidth,angle=90]{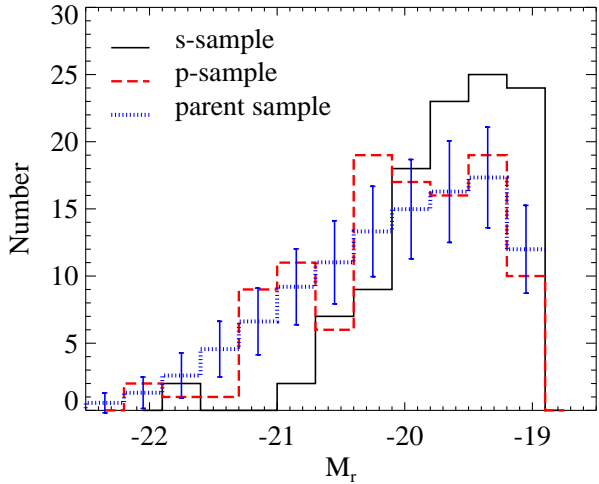}
\caption{The distribution of r-band absolute magnitude of void galaxies. 
The solid and dashed lines present the distributions for sparse and populous 
samples, respectively. 
 The dotted line presents the distribution for our parent sample obtained by 
a Monte Carlo simulation and the errorbars are the 
standard deviation of each bin (see text for description). }
\label{absolute}
\end{figure}

Absolute luminosity is a fine tracer of the total mass of galaxies.
Hence, to study the distribution  of masses of void galaxies we use their
 luminosity as a proxy.
Unlike over-dense regions, it is expected that the probability of
finding massive halos in void region to be small. 
Lack of merger events can be a logical
explanation of such observations.

Fig. \ref{absolute} presents the distribution of the r-band absolute magnitude 
of void galaxies measured from Petrosian
magnitude \citep[][]{Petrosian1976}. 
The s- and p-samples are drawn using solid and dashed
lines, respectively. As it can be seen galaxies in the s-sample
have a distribution peaked at $\sim$ -19.5 with few galaxies brighter
than $\sim$ -21. On the contrary, the p-sample shows a broader distribution that
extends to $\sim$  -22. Therefore, voids of higher density contrasts
can host significantly brighter (presumably more massive) galaxies than those of
lower density counterpart. Using a  Kolmogorov-Smirnov (KS) test we find that the
probability of the two samples 
to be drawn from the same parent distribution is negligible (zero). Hence, 
the difference between the two samples reported here is statistically significant.
We further check how different are these two samples in comparison with our  
parent sample (39750 galaxies) from which we have extracted our void catalog. 
This exercise will demonstrate how the void galaxy luminosity distribution may 
differ from the luminosity distribution of galaxies across the local universe 
as a whole.
To do so, we try a Monte Carlo analysis as following: (1) randomly choose 110 galaxies out 
of 39750 (2) calculating number of galaxies in each magnitude bin (3) repeating  steps (1) and (2)  
1000 times (4) and finally finding the mean and standard deviation of the 1000 numbers in each bin.
We have chosen 110 galaxies at step (1), to keep the same number of galaxies as those of s- or p-sample. 
The mean and standard deviation obtained in each bin are shown as 
dashed-dotted histogram and errorbars in Fig. \ref{absolute}. 
This analysis shows that the p-sample closely follows the parent distribution. Running a KS test we find more than 
80\% probability that parent and p-sample to have the same distribution. In stark contrast  parent and s-sample 
present a very different magnitude distributions with a zero percent KS test probability.

Existence of massive object in 'p-sample',
might be due to the hierarchical nature
of structure formation and/or high efficiency of star formation in their progenitors.
This indicates that the formation of
void galaxies and their path of evolution can strongly depend on their
environmental properties.  Discriminating between galaxies in various voids,
our results also provide an interesting tool to test the prediction of
cosmological dark matter simulations and semi-analytical models.

\begin{figure}
\centering
\hspace{-1.00cm}\includegraphics[width=0.40\textwidth,angle=90]{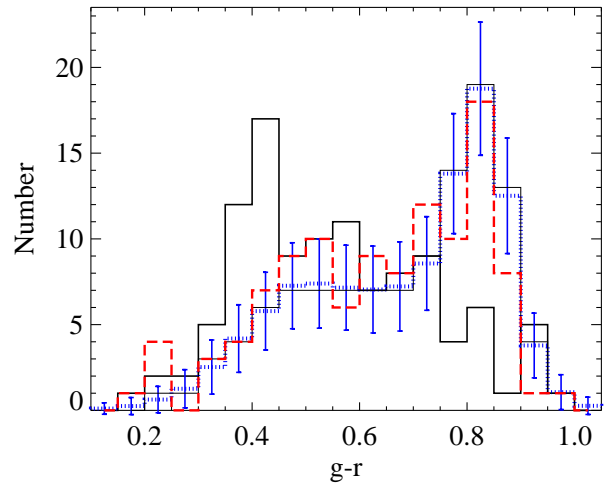}
\caption{The g-r color distribution of sparse and populous  void
galaxies are shown in solid and dashed lines, respectively. The K-S
test shows that the probability of two samples being drawn from
similar distribution is only about 0.002. 
 The dotted line presents the distribution for our parent sample obtained by 
a Monte Carlo simulation and the errorbars are the 
standard deviation of each bin (see text for description). 
}
 \label{color}
\end{figure}

\subsection{Color}

The color of galaxies can be used to probe     
their dominant stellar populations and star formation
history.
Generally bluer galaxies have younger stellar population 
in comparison with red galaxies. 
It is also known that blue galaxies are dominated by
late types while the red galaxies are dominated by early types \citep[e.g.][]{Strateva2001}.

Within the hierarchical framework of $\Lambda$CDM, galaxies assemble their masses over time via
different modes.
Depending on the physical processes and when they act on shaping the galaxy, the resulting stellar 
populations can become redder or remain blue through sustained star formation. Since processes like 
merging and gas accretion are important, environment can strongly regulate the evolution of 
galaxies.
This picture demonstrates why galaxy environment appears to play a key
role in controlling the stellar population
properties of the galaxies and they are the product of a complex assembly and environment history.
Observations of void galaxies selected by different samples show that
statistically they are gas rich, blue and late-type disk galaxies 
\citep[][]{Rojas2004,Rojas2005,Patiri2006,kreckel2011b,Kreckel2012}.

Here,we investigate the color differences between
void galaxies in the sparse and populous samples.This approach allows
us to see how galaxy color depends on properties of host voids, $\delta$.
To do so, we use the model color ($g-r$) which is derived from the SDSS model
magnitudes. For each galaxy, these are derived from the best fitting
de Vaucouleur \citep{deVaucouleur1948} or exponential profiles \citep[][]{Freeman1970}.
Fig. \ref{color} shows the color distribution of void galaxies in the range of ${\sim}$ 0.2 -- 0.9.
Although both distributions have a wide range of colors, 
a bimodality is clearly visible. 
The s- and p-sample present single peak around $g-r=0.4$ (blue) and 0.8 (red), respectively.
Repeating the same Monte Carlo analysis as that in section \ref{luminosity} we find a 
30\% probability that parent and p-sample to be drawn from the same distribution (dotted histogram in Fig. \ref{color}). 

Although the star formation history of a galaxy is a function of stellar mass,
the $g-r$ distribution of void galaxies might include a real
evolutionary effect, caused by the dependence of the red and blue void galaxy on the density contrast
of a void. In other words, regions with different initial cosmological density fields might result
in different galaxy populations,namely {\it{active}} or {\it{passive}}.

\begin{figure*}
\center{
\hbox{
\hspace{-1.0cm}
\includegraphics[width=0.40\textwidth,angle =90]{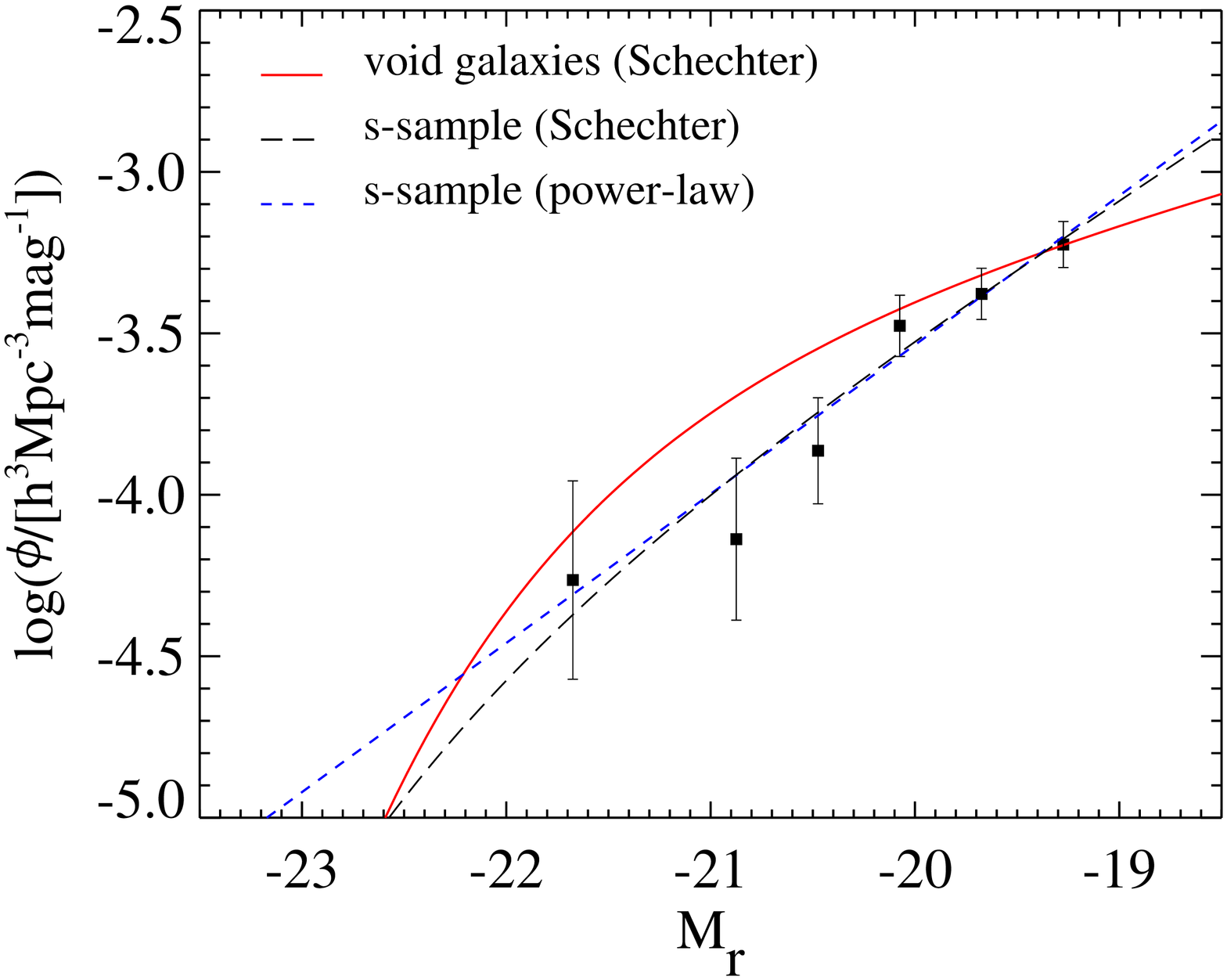}
\includegraphics[width=0.40\textwidth,angle =90]{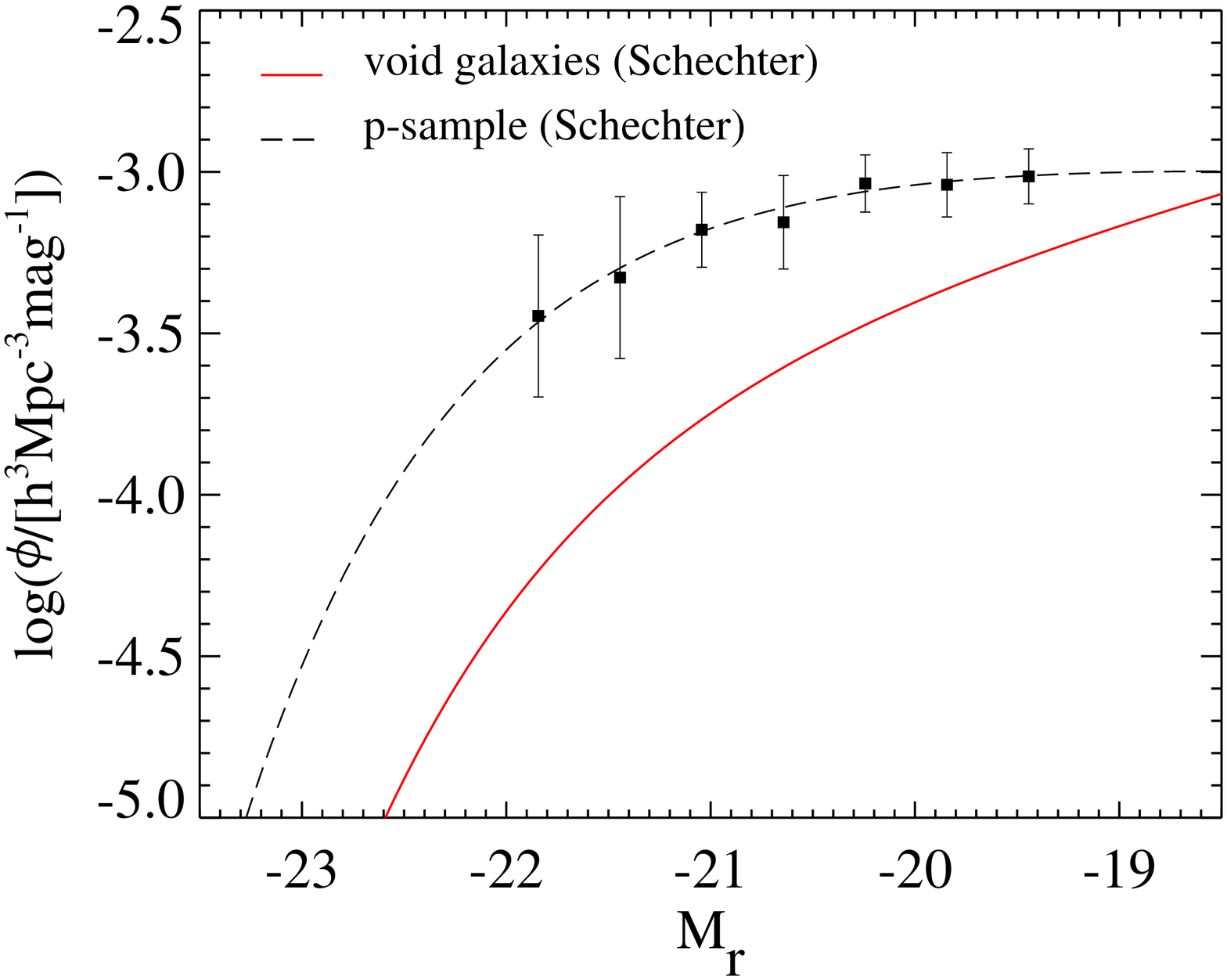}
}
\caption {{\it Left:} the LF of s-sample. The long-dashed and dashed lines 
present the best fitted Schechter and power-law, respectively.    
{\it Right:} the LF of p-sample. The long-dashed line presents the best fitted 
Schechter. 
The solid line in both panels shows the best fitted Schechter to all 1014 void galaxies. 
The p-sample galaxies are fitted with a Schechter function while the LF of the s-sample 
appears to be following a power-law.}
\label{lf}}
\end{figure*}

\subsection{Luminosity Function}\label{luminosity}
\begin{table}
\begin{center}
\caption{Schechter fitted parameters for different sample of void galaxies.}
\begin{tabular}{ l c c }
\hline
samples & $\alpha$ & $M^{*}$\\
\hline
  all void galaxies & -1.4$\pm$0.2 & -21.4$\pm$0.5 \\
  s-sample          & -2.0$\pm$0.5 & -22.5$\pm$4.1 \\
  p-sample          & -0.9$\pm$0.5 & -21.5$\pm$1.3 \\
\hline
\end{tabular}
\label{tablf}
\end{center}
\end{table}
One of the key statistical tools to study the galaxy distribution is the luminosity function (LF).
One can describe the global properties of galaxy populations
and study the  formation and evolution of galaxies through the LF. To understand how galaxies form, we also need to
understand how the LF depends on the environment. The influence of the local environment
 on the LF from over-dense to under-dense regions (supercluster/void) has been
investigated by several authors %
\citep[e.g.][]{Barkhouse2007,Bai2009,Robotham2010,Zandivarez2011}.
Although there are many LF studies using different samples and
approaches at different redshifts \citep[e.g.][]{Johnston2011}, the majority of them are related to galaxies in
over-dense regions. Not many have explored the LF of void galaxies \citep{Hoyle2005}.
It is not yet clear how the LF of void galaxies depends on the properties of their host void.

Here we use s- and p-sample to study the LF of void galaxies in
 different voids, taking the effect of density contrast into account.
In Fig. \ref{lf} we show the LF of s- and p-sample in the r-band
Petrosian magnitude as squared symbols with error bars.
We describe the LFs using Schechter function \citealt{Schechter1976}, which has the following shape
\begin {equation}
\phi(L)=\phi^{*} (L/L^{*})^{\alpha} \exp-(L/L^{*})
\end{equation}
where $\alpha$ , $L^{*}$ and $\phi^{*}$ are the three parameters to fit.
The best fitting LF for different samples are shown in Fig. \ref{lf} where the parameters
are given in Table \ref{tablf}.
Clearly the LF of s-sample does not follow a Schechter function.
This can also be inferred from the large errors in the $M^{*}$
parameters. 
The large error in the $M^{*}$ indicate the insensitivity of the LF of the s-sample to this parameters. Further, 
because the fit passes through 1-$\sigma$ of all points, the LF of this sample follows a power-law. Fitting a pure 
power-law we find a power-law index of $\alpha$ = $-2.15\pm0.21$ with a reduced  $\chi_\nu$= $0.54$ and. While the $\chi_\nu$ of power-law 
is smaller than that of Schechter ($\chi_\nu$= $0.70$) but the power index of both are consistent within the errors.
The LF of all void
galaxies as well as that of p-sample are well fitted with the
Schechter. The relatively large error in $M^{*}$ of the p-sample is
due to the large errors in its LF which caused by the number
statistics.

There are clear differences between the  LF of s- and p-sample which is mainly
due to the lack of bright galaxies in the sparse sample. Moreover, while the LF of the p-sample
follows a Schechter function, it seems like a power-law for the s-sample. Gaussian and double Schechter
have been alternatively used to describe the LF of galaxies. Even a cursory look at the  LF of s-sample
shows a Gaussian would not fit it. Further, fitting a double Schechter which has 6 parameters
in the current LF does not seem to be statistically reasonable. Hence, the available data
does not allow us to further investigate it.
the detailed differences in the shapes of the LF for the two void galaxy samples implies that 
the possible variety of formation and/or evolution mechanisms are a function of galaxy density 
even among obvious voids.

\section{Discussion}
\label{discussion}

In this paper we have studied the photometric properties of void galaxies based on void catalog on SDSS DR10 at $z=0.010 - 0.055$. Our
void catalog consists of a large variety of voids from small to large and encompasses a range in density contrast from
low to high population. In order to investigate how the density contrast of voids, affects the evolution of void galaxies, we define two subsample
of void galaxies which are located in sparse and populous  voids. Our results indicate that the two populations show systematic differences in
photometric properties such as luminosity, color distribution and the luminosity function.
 While the luminosity distribution of galaxies in populous voids follows the luminosity 
distribution of the general population of galaxies in SDSS within $0.010 < z < 0.055$, the 
luminosity distribution of galaxies in sparse voids show that they are generally dimmer. 
Also the colors of galaxies residing in sparse voids are bluer 
and the galaxy generally less luminous indicating that they are likely to have low 
but sustained rates of inefficient star formation throughout their evolution. 

Furthermore, the LF of 
galaxies in sparse voids do not follow a Schechter function, seen in the populous void galaxies. 
In this letter we have shown clear indications that the voids with different density 
contrasts also host different galaxy populations.What is also quite interesting is the 
similarity between the properties (luminosity, color and LF) of galaxies in populous voids 
and the general population of galaxies in the local universe. It is not unimaginable that 
sparse voids could be the least evolved voids in the context of hierarchical structure 
formation \citep{Sheth2004}. 
Based on this indicative study, one could argue that populous voids 
contain a mixed population of galaxies which might be the consequence of mergers among voids, contrary 
to the sparse voids which seem to present a more homogeneous galaxy population.  

The purpose of this study was to highlight the important role of the 
density contrast, specially at the extreme low density environments of the voids, s-sample. 
Now, having shown that the properties of galaxies depend on whether or not a void 
is sparse or populous in a non-trivial way, it is important to determine why some voids 
are sparse and some are populous to truely understand how environmental density 
affects galaxy evolution and what processes regulate this evolution.   
\section*{Acknowledgement}
We would like to thank Roya Mohayaee 
and Gary Mamon for useful 
discussions.We also thank the referee for his constructive
comments.

\end{document}